\documentclass[twocolumn,amsmath, amssymb, superscriptaddress, aps, prd, 10pt, longbibliography, showkeys]{revtex4-2}

\usepackage{xcolor, cancel, comment, amsmath, lipsum, orcidlink, hyperref}
\hypersetup{
 pdftitle={ASTRAL}, % title
 pdfauthor={Gravely}, % author
 colorlinks=true,
 citecolor={blue},
}

\newcommand{\ab}[1]{\left|#1\right|}

\newcommand{\pa}[1]{\left(#1\right)}

\newcommand\authors[1]{{\color{black}{#1}}}

\begin{document}

\title{ASTRAL: A Framework for Optimal Placement and Emulation of Stellar Evolution Models}

\author{Trevor Gravely\,\orcidlink{0009-0004-6058-843X}}
\affiliation{Department of Physics \& Astronomy, Vanderbilt University, Nashville TN 37212, USA}
\affiliation{Center for Theoretical Astrophysics, Los Alamos National Laboratory, Los Alamos, NM 87545, USA}

\author{Marko Risti\'{c}\,\orcidlink{orcid=0000-0001-7042-4472}}
\altaffiliation{ISTI Fellow}
\email{mristic@lanl.gov}
\affiliation{Center for Theoretical Astrophysics, Los Alamos National Laboratory, Los Alamos, NM 87545, USA}
\affiliation{Theoretical Division, Los Alamos National Laboratory, Los Alamos NM 87545, USA}

\author{Brandon Barker\,\orcidlink{0000-0002-8825-0893}}
\altaffiliation{Metropolis Fellow}
\affiliation{Center for Theoretical Astrophysics, Los Alamos National Laboratory, Los Alamos, NM 87545, USA}
\affiliation{Computational Division, Los Alamos National Laboratory, Los Alamos NM 87545, USA}

\date{\today}

\begin{abstract}

Generating dense grids of detailed stellar evolution simulations is computationally prohibitive, as resolving progressively complex nucleosynthesis chains and dynamical timescales incurs a rapidly increasing computational cost that scales with dimensionality and resolution.
Existing simulation grids are sparse, heterogeneous, and often cover disparate regions of parameter space, limiting (direct) systematic comparison and broader generalization to downstream science analyses. We present ASTRAL (\textbf{A}ctive \textbf{S}imulation \textbf{T}uning and \textbf{R}egression for \textbf{A}strophysical \textbf{L}ibraries), a publicly available framework for optimal placement and emulation of stellar evolution models. The ASTRAL framework builds surrogate models that provide a continuous representation of the parameter space which is populated by sparse simulation grids. Furthermore, ASTRAL interpolates across the simulation parameter space (e.g. zero-age-main-sequence mass) to predict stellar properties of stars with associated uncertainty estimates for systems not explicitly simulated. We show that ASTRAL significantly outperforms uniform placement in constructing emulation-ready model libraries that capture both global and local variability in the output features. We showcase the performance of our framework in the context of a simple, one-dimensional MESA model. Finally, we examine how our methodology can inform downstream applications by analyzing our emulator’s predictive confidence when applied to supernova simulations. We emphasize that, while developed for stellar evolution, the ASTRAL framework is directly extensible to other simulation-driven domains where model evaluations are expensive and parameter spaces are high-dimensional.
\end{abstract}


\maketitle

\section{Introduction}
\label{sec:intro}
% =============================================================================

Stellar evolution modeling underpins our understanding of stars across cosmic time, shaping phenomena ranging from star formation to the final pre-supernova structure that gives rise to many of the transient events observed in the Universe \citep{kippenhahn2012, weiss:2004, abbott:2016, abbott:2017a, cahillane:2022, abbott:2023}.
Uncertainties in stellar evolution modeling propagate directly into predictions of supernova outcomes, compact object populations, and other downstream observables, highlighting the need for deliberate uncertainty mitigation. 
One approach to systematic uncertainty reduction favors understanding populations of
massive stars as opposed to focusing on highly specific, idealized case studies. Understanding the distribution of phenomena dictated by populations of massive stars requires a collection of stellar evolution models that are equally as diverse as those that may be realized in nature.
In practice, such a collection requires sampling the full diversity of stellar evolution outcomes across parameter space using dense, expansive, 
multidimensional grids of massive stars evolved out to late times, spanning zero-age-main-sequence (ZAMS) mass, metallicity, rotation rates, and more complex properties like binarity \citep{2012Sci...337..444S, 2021A&A...656A..58L, 2025A&A...695A..71L}.
Even in simplified one-dimensional models, stellar evolution simulations \citep[e.g., KEPLER and MESA;][]{weaver:1978, paxton:2011, paxton:2013, paxton:2015, paxton:2018, paxton:2019, jermyn:2023} are computationally expensive and exhibit numerical instability, thus inhibiting long term evolution and complete grid construction. As a star evolves beyond core helium burning, the nuclear reactions responsible for core burning become increasingly stiff, requiring more demanding numerical methods with increasingly small time steps. The increasing number and complexity of nuclear reactions and the added complications introduced by neutrino losses act as significant bottlenecks in continuing the evolution of the star to late times \citep[see discussion in Ref.][]{grichener2025}. 
These instabilities and computational costs inhibit the construction of dense, detailed stellar evolution grids which are suitable for astrophysical inference (although prior efforts have tackled the issue, e.g. \cite{2014A&A...570A..66S}), necessitating alternative techniques for broad parameter space representation and grid construction.

Standard fitting formulae have historically been applied to understanding stellar and binary systems \citep{2000MNRAS.315..543H, 2020MNRAS.495.4170T}, with recent studies also leveraging now established machine learning approaches, in particular Gaussian processes \citep{2017IAUS..325...46B, 2018PhRvD..98h3017T}. Gaussian processes (GPs) provide a powerful framework for emulating expensive simulations and quantifying predictive uncertainty \citep{Rasmussen_2006} and have been implemented across many astrophysical applications including gravitational waveform surrogate modeling \citep{Doctor17}, kilonova and supernova light curve emulation \citep{Coughlin18, Ristic22, sarin:2025}, and galaxy profile fitting \citep{Liu25} (see Ref. \cite{aigrain_gaussian_2022} for a more exhaustive list of examples). Previous studies in the literature have explored scalable stellar evolution parameter forecasting using both deep learning \citep{2024A&A...681A..86M} and Gaussian processes \citep{2022MNRAS.511.5597L}. In a similar vein, population synthesis studies apply machine learning methods to interpolate and emulate stellar profiles across binary evolution model grids \citep{2025ApJS..281....3A, 2025A&C....5100935T}. 

Recent work has further leveraged GP predictive uncertainties (and other machine learning tools, e.g. \cite{2022ApJ...938...64R}) to guide simulation placement, enabling adaptive sampling strategies that prioritize regions of high uncertainty to achieve global uncertainty reduction \citep{Ristic22, Ristic_lGRBKNe}. When combined, these strategies establish a powerful emulation setup: simulation optimization ensures that each placed simulation provides maximal information, while the GP-derived placement metric all but ensures convergence and uncertainty minimization for the emulation step. This synergy is particularly appealing in contexts like complex stellar evolution models, where computational expense and the need for a low-cost representation of the parameter space beget emulation. 

In this work, we present ASTRAL, a new simulation placement and emulation
framework that leverages Gaussian process regression to interpolate and
predict stellar evolution model outputs. ASTRAL uses GP regression to interpolate an
existing grid of stellar evolution models, and produces predictions of key
stellar evolution output parameters as a function of standard stellar inputs, in
this case focused solely on the zero-age main sequence mass.
Additionally, we show that ASTRAL is capable of optimizing the placement of
computationally expensive simulations while significantly outperforming standard placement strategies.

This paper is organized as follows. In
Section~\ref{sec:methods} we describe our methodology, specifically our simulation assumptions, 
a brief overview of the GP regression technique, and our application of GPs to uncertainty-informed simulation placement.
In Section~\ref{sec:results} we present our results which focus on GP interpolation of stellar evolution outputs,
simulation grid building through optimized parameter placement, and the propagation of uncertainty minimization between
stellar evolution and supernova simulation outputs. We present an overarching discussion of our results in Section~\ref{sec:disc} and 
end with our conclusions in Section~\ref{sec:conc}.

\section{Methodology}
\label{sec:methods}

At a high level, our methodology proceeds as follows. We establish a reference for interpolation and regression using the pre-existing dataset from Ref. \cite{Sukhbold_2016} and a standard Python library by generating scalar GP emulators of the stellar evolution output parameters as a function of zero-age main sequence mass ($M_{\rm{ZAMS}}$). Expanding on this reference, we present a metric which guides the reduction of GP prediction variance through an optimized simulation placement approach. We evaluate the effectiveness of this approach compared to standard (uniform) simulation placement using a simple MESA stellar evolution model. Finally, we revisit the initial stellar evolution dataset \cite{Sukhbold_2016} in the context of the follow-up supernova simulations informed by that grid as presented in Ref. \cite{barker:2022}. We evaluate our simulation placement metric's effectiveness on reducing GP prediction variance of supernova simulation outputs when trained on upstream stellar evolution simulations.

\subsection{Stellar Evolution and Supernova Simulations}
\label{sec:sesn}
We use the dataset from Ref. \cite{Sukhbold_2016} generated using the KEPLER code \citep{Weaver_1978}, the physics of which are discussed in e.g., Ref. \cite{Woosley_2002}. We will refer to this dataset as the ``KEPLER grid" throughout the rest of this text. In this study, we examine only the models for which $9 \leq M_{\rm{ZAMS}} \leq 30$ to focus on stellar evolution models with retained hydrogen envelopes (for the supernova connection described at the end of this section). This amounts to 187 stellar models. All stars are non-rotating with initial solar composition and evolve from the zero-age main sequence until iron-core collapse. Each simulation yields numerous scalar output parameters as a function of $M_{\text{ZAMS}}$, several of which we consider: the pre-supernova mass, radius, temperature, and luminosity; the central entropy, electron fraction, and density; compactness parameters $\xi_{M}$ \citep[][]{oconnor:2011} defined via
\begin{equation}
    \xi_M = \frac{M / M_{\odot}}{R(M) / 1000\ \text{km}},
\end{equation}
and the mass of the He/CO/Si/Fe cores and their corresponding radii. 
Additional details about the KEPLER stellar evolution simulation data we use can be found in Section 2.1 of Ref. \cite{Sukhbold_2016}.

The KEPLER models are used as inputs for subsequent simulations in spherical symmetry through collapse and potential explosion using the neutrino-driven, turbulence aided method STIR \citep{couch2020} in the FLASH code \citep{fryxell2000, dubey2009, dubey2022}. STIR includes multigroup two moment ``M1'' neutrino radiation transport, a dense matter equation of state \citep{steiner:2013}, a general relativistic effective potential, and a Reynolds averaged Navier-Stokes treatment of turbulence and convection in spherical symmetry. These explosions are followed through shock breakout producing electromagnetic observables \citep{barker:2022} using SNEC \citep{morozova:2015}. Thus, we obtain a consistent, corresponding grid of supernova outputs from the upstream KEPLER grid. We will refer to the supernova simulation library as the ``KEPLER-FLASH grid'' throughout the text.

% MESA
We demonstrate the capability to guide stellar evolutionary modeling \emph{in-situ}
by running stellar evolution simulations using MESA \citep[version r24.08.1;][]{paxton:2011, paxton:2013, paxton:2015, paxton:2018, paxton:2019, jermyn:2023}. We use a basic stellar evolution model to reduce the total build time for the proof-of-concept grid --- specifically, we use the \texttt{12M\_pre\_ms\_to\_core\_collapse} test within the MESA test suite, which evolves a solar metallicity model from the pre-main sequence to the onset of core collapse. We note that although we opted to use MESA due to its role as an open-source software the results presented here apply to any stellar evolutionary modeling framework and beyond.

% GP General

\subsection{Gaussian Processes}
\label{sec:gp}

Below, we present a high-level overview of GP regression. We defer readers to Ref. \cite{Rasmussen_2006} for an exhaustive general review of GP regression and Ref. \cite{aigrain_gaussian_2022} for a more field-specific review of GP regression using astronomical data.

Let $\{ y(x) : x \in X \}$ be a stochastic process, where each $y(x)$ is a random variable and $x \in X$ is an independent variable for some domain $X$. A stochastic process is a Gaussian process if the distribution of any finite set of random variables in the process is jointly Gaussian 
\begin{align}
    \label{eq:joint-gaussian-pdf}
    p(y(x_1), ..., y(x_n)) &= \frac{1}{\sqrt{2 \pi \ab{\mathbf{K}}}} \nonumber \\
    & \times \exp \pa{ 
        - \frac{1}{2} (\mathbf{y} - \mathbf{m})^T \mathbf{K}^{-1} (\mathbf{y} - \mathbf{m})
    },
\end{align}
where each $x_i \in X$ and $\mathbf{y} \equiv (y(x_1), ..., y(x_n))$. Here, $\mathbf{m} \in \mathbb{R}^n, \mathbf{K} \in \mathbb{R}^{n \times n}$ are the mean and covariance of the GP respectively --- specifying a particular GP amounts to identifying a particular choice of mappings from $(x_1, ..., x_n) \mapsto \mathbf{m}, \mathbf{K}$.

In particular, in GP regression one assumes that the covariance matrix takes the form
\begin{equation}
    \label{eq:covariance-matrix-form}
    (\mathbf{K})_{ij} = k(x_i, x_j ; \boldsymbol{\phi}) + \delta_{ij} \sigma_n^2.
\end{equation}
Here $\boldsymbol{\phi}$ is a set of model parameters that parameterize the functional form of $k$, the kernel function, and $\sigma_n$ is a parameter characterizing the noise scale given a particular application. Equation \eqref{eq:covariance-matrix-form} contains two contributions: first, the intrinsic observational noise, which is quantified via the term $\delta_{ij} \sigma_n^2$, only contributing along the diagonal of $\mathbf{K}$; and second, the correlations between inputs, which are quantified via the kernel function $k$. The choice of $k$ is free and often dictated by what is best suited to the problem at hand. In this work, we assume a radial basis function (RBF) kernel
\begin{equation}
    \label{eq:covariance-kernel}
    k(x_i, x_j ; \boldsymbol{\phi}) = \alpha^2 \exp \pa{ - \frac{\ab{x_i - x_j}^2}{2 \ell^2} },
\end{equation}
where the kernel parameters $\boldsymbol{\phi} = (\alpha, \ell)$ control the overall scale of fluctuations and the correlation length, respectively. \authors{For the GP presented in this work, we find no preferential performance between the RBF and Matern 3/2 and Matern 5/2 kernels. However, GP behavior may be more sensitive to kernel choices for more complex problem setups, especially those with coupled input dimensions, and thus we encourage readers to treat the choice of kernel as a sensitive hyperparameter until proven otherwise.}  

Equation \eqref{eq:joint-gaussian-pdf} also contains a mean $\mathbf{m} \in \mathbb{R}^n$ which one assumes to be given by
\begin{equation}
    (\mathbf{m})_{i} = m(x_i, \boldsymbol{\theta}),
\end{equation}
for some mean function $m$ and some model parameters $\boldsymbol{\theta}$. Without loss of generality, the mean can be absorbed into the kernel by demeaning the data on which one regresses. We correspondingly set $m(x, \boldsymbol{\theta}) \equiv 0$, leaving $\{ \boldsymbol{\phi}, \sigma_n \}$ as the only parameters in the model.

To fit the model to some set of data $\mathbf{y} \in \mathbb{R}^n$ corresponding to inputs $\mathbf{x} = \pa{ x_1, ..., x_n }$, we maximize the likelihood
\begin{equation}
    \mathcal{L}(\boldsymbol{\phi}) = p(\mathbf{y} | \boldsymbol{\phi}, \sigma_n),
\end{equation}
where $p(\mathbf{y} | \boldsymbol{\phi}, \sigma_n)$ is given by Equation \eqref{eq:joint-gaussian-pdf} with inputs $\mathbf{x}$. 

Once the GP model parameters have been fit to the data, the GP can be used to predict means and variances of the underlying stochastic process at $n_*$ unseen locations. In particular, if $\mathbf{y}$ are existing observations corresponding to inputs $\mathbf{x}$, the predictions $\mathbf{y}_* \in \mathbb{R}^{n_*}$ at unseen inputs $\mathbf{x}_* \in X^{n_*}$ are distributed as a joint Gaussian with mean $\mathbf{f}_* \in \mathbb{R}^{n_*}$ and covariance $\mathbf{C}_* \in \mathbb{R}^{n_* \times n_*}$ given by 
\begin{gather}
    \label{eq:predictive-mean-and-covariance}
    \mathbf{f}_* = \mathbf{m}_* + \mathbf{K}_*^T \mathbf{K}^{-1} (\mathbf{y} - \mathbf{m}), \\ 
    \mathbf{C}_* = \mathbf{K}_{**} - \mathbf{K}_*^T \mathbf{K}^{-1} \mathbf{K}_* + \sigma_n^2 \mathbf{Id},
\end{gather}
where $\mathbf{m}_* \equiv m(\mathbf{x}_*, \boldsymbol{\theta}) \in \mathbb{R}^{n_*}$ and 
\begin{gather}
    \label{eq:predictive-covariance-matrices}
    (\mathbf{K}_*)_{ij} = k(x_{i}, x_{*j}; \boldsymbol{\phi}) \in \mathbb{R}^{n \times n_*}, \\
    (\mathbf{K}_{**})_{ij} = k(x_{*i}, x_{*j}; \boldsymbol{\phi}) \in \mathbb{R}^{n_* \times n_*}.
\end{gather}
We keep $\mathbf{m}$ in Equation \eqref{eq:predictive-mean-and-covariance} for completeness, but in all models considered in this work we take $m(x, \boldsymbol{\theta}) \equiv 0$ for all $x \in X$. Also note that with the joint distribution in hand, one can marginalize over observations to obtain the variance in a single predicted observation.

% GP Kepler

\subsection{Gaussian Processes Applied to KEPLER}
\label{sec:gp_kepler}

We begin by training a separate scalar GP on each of the 22 KEPLER stellar evolution outputs described in Section \ref{sec:sesn}. While we do expect correlation between the KEPLER outputs, our goal of identifying an optimal placement metric benefits from separate predictions; we are interested in choosing a parameter that both varies smoothly as a function of $M_{\rm{ZAMS}}$ and is physically well motivated. We normalize (z-score) both the input masses $M_{\rm{ZAMS}}$ and the KEPLER output parameters (separately) for GP training. We implement Gaussian process regression using \texttt{scikit-learn} \citep{pedregosa2018scikitlearnmachinelearningpython}, specifically the \texttt{GaussianProcessRegressor} module. Our kernel matches the functional form outlined in Equation~\eqref{eq:covariance-kernel} and assumes all the default settings. We optimize hyperparameters using the default \texttt{scikit-learn} Gaussian process optimizer, \texttt{fmin\_l\_bfgs\_b}, a limited-memory BFGS algorithm. 

\subsection{Gaussian Processes Applied to MESA}
\label{sec:gp_mesa}

% GP MESA

We extend the use case of our GPs to inform simulation placement in $M_{\text{ZAMS}}$ during the construction of uncertainty-informed grids of stellar evolution models simulated with MESA. We choose $M_{\rm{Fe}}$, the iron core mass, as the output parameter which guides our uncertainty-informed simulation placements as it is a key parameter in determining the final remnant masses and explosion mechanisms relevant to downstream supernova simulations, displays non-monotonicity and stochasticity, and is a core output parameter in the KEPLER grid, providing cohesion with Section~\ref{sec:gp_kepler}. While we employ only a single placement parameter for simplicity, an uncertainty metric $u$ can combine any number of output parameters by taking the form
\begin{equation}
    u(M_{\text{ZAMS}}) = \sum_{i = 1}^{N} \alpha_i \sigma_i(M_{\text{ZAMS}}),
\end{equation}
where $\sum_i \alpha_i = 1$ for $N$ output parameters, each with GP predictive uncertainty $\sigma_i$. The weights $\alpha_i$ control the contribution of each output parameter to the simulation placement process, thus allowing domain-specific tailoring for downstream applications that might be more sensitive to particular parameters. In this work, $u$ and $\sigma_i$ only depend on $M_{\text{ZAMS}}$, but will generally be functions of the full range of input parameters varied in the stellar evolution models.

We define the total uncertainty $U$ of a given GP as the average thickness of the $95\%$ confidence band over the dynamic range of the training set 
\begin{equation}
    \label{eq:total-uncertainty}
    U = \frac{1}{x_{\text{max}} - x_{\text{min}}} \int_{x_{\text{min}}}^{x_{\text{max}}} \pa{ y_{u}(x) - y_{\ell}(x) }\ dx,
\end{equation}
where $y_{u, \ell}(x)$ are the $95\%$ GP confidence bounds and $x_{\text{min / max}}$ are the smallest and largest $M_{\rm{ZAMS}}$ samples in our grid. The baseline grid to which we compare against is a uniform grid across a fixed dynamic range of $9\ M_\odot \leq M_{\text{ZAMS}} \leq 30\ M_\odot$ with $65$ equally spaced simulations. The grid building procedure is as follows:
\begin{enumerate}
    \item Construct a uniform initial training grid of 17 simulations for $9\ M_\odot \leq M_{\rm{ZAMS}} \leq 30\ M_\odot$.
    \item Train the Gaussian process on the current grid iteration. 
    \item Determine $M_{\text{ZAMS}}^*$, the $M_{\text{ZAMS}}$ with maximal predictive uncertainty for the output parameter(s) of interest (here $M_{\rm{Fe}}$).
    \item Place a new MESA simulation at $M_{\text{ZAMS}}^*$.
    \item Repeat Step 2 until the grid reaches the fixed size of the baseline grid (65 simulations).
\end{enumerate}

We note that the choice of 17 initial simulations is arbitrary; different choices of the initial number of simulations will yield varying results. Generally, however, more simulations in the initial grid will contribute to quicker convergence toward a desired GP variance (though performance will vary depending on the complexity of the input parameter space).

\subsection{Gaussian Processes Applied to KEPLER-FLASH}
\label{sec:gp_flash}

In Section~\ref{sec:gp_mesa}, we describe how GP variance influences the placement and generation of MESA simulations, assuming that additional simulations can be readily generated by the user. We slightly adapt our simulation placement metric to investigate how uncertainty-informed fixed-grid sampling of KEPLER outputs affects downstream supernova models. We note that not all KEPLER stellar evolution models result in supernovae in the STIR model; black hole formation is also possible. Therefore, we first down-select the KEPLER grid based on which models lead to explosion. We are then left with two separate simulation grids: 136 down-selected stellar evolution models (the KEPLER grid) and the 136 supernova models derived from these stellar evolution models (the ``KEPLER-FLASH" grid). We emphasize that the KEPLER-FLASH grid of stellar-evolution-into-supernova models \emph{is distinct from the KEPLER grid of only stellar evolution models discussed in Section \ref{sec:sesn}}. Our goal is to examine how training GPs on the KEPLER grid affects the GP prediction variance for the KEPLER-FLASH grid parameters. 

We carry out the following modified procedure:

\begin{enumerate}
    \item Initialize a pseudo-uniform grid of simulations by sequentially selecting every third $M_{\text{ZAMS}}$ model from the KEPLER grid
    \item Train the Gaussian process on the current grid. 
    \item From the remaining stellar evolution models in the KEPLER grid, determine $M_{\text{ZAMS}}^*$, the $M_{\text{ZAMS}}$ with maximal predictive uncertainty for the output parameter(s) of interest (here $M_{\rm{Fe}}$). 
    \item Add this model to the constructed grid.
    \item Repeat Step 2 until 35 simulations have been added to the grid.
\end{enumerate}

Our initial simulation grid inherits pseudo-uniformity from the KEPLER grid, which is piece-wise uniform in $M_{\rm{ZAMS}}$. We ultimately obtain a constructed supernova grid of $45 \text{ (initial grid)} + 35 \text{ (uncertainty-placed models)} = 80$ models out of the total 136 from the KEPLER-FLASH grid. For comparison, we also construct a baseline grid of $80$ models by randomly selecting $35$ models from the KEPLER-FLASH grid and adding them to the initial grid. In both cases, we are using only $\sim 60 \%$ of the simulations in the full KEPLER-FLASH grid. 

\section{Results}
\label{sec:results}

Here we present our results for the three separate analyses described in Section~\ref{sec:methods}. As a reminder, we first examine the GP fits to stellar evolution output parameters as a function of $M_{\rm{ZAMS}}$ as described in Section~\ref{sec:gp_kepler}. We then analyze GP uncertainty across a library of simple MESA stellar evolution models, comparing grids constructed by optimizing over GP uncertainty versus uniform sampling, as described in Section~\ref{sec:gp_mesa}. Finally, we compare the performance of GP predictions of supernova simulation outputs using a subset of models chosen through uncertainty minimization of stellar evolution outputs as described in Section~\ref{sec:gp_flash}.

\subsection{GP Fits to KEPLER Simulation Outputs}

\begin{figure}[ht!]
    \includegraphics[width=\linewidth]{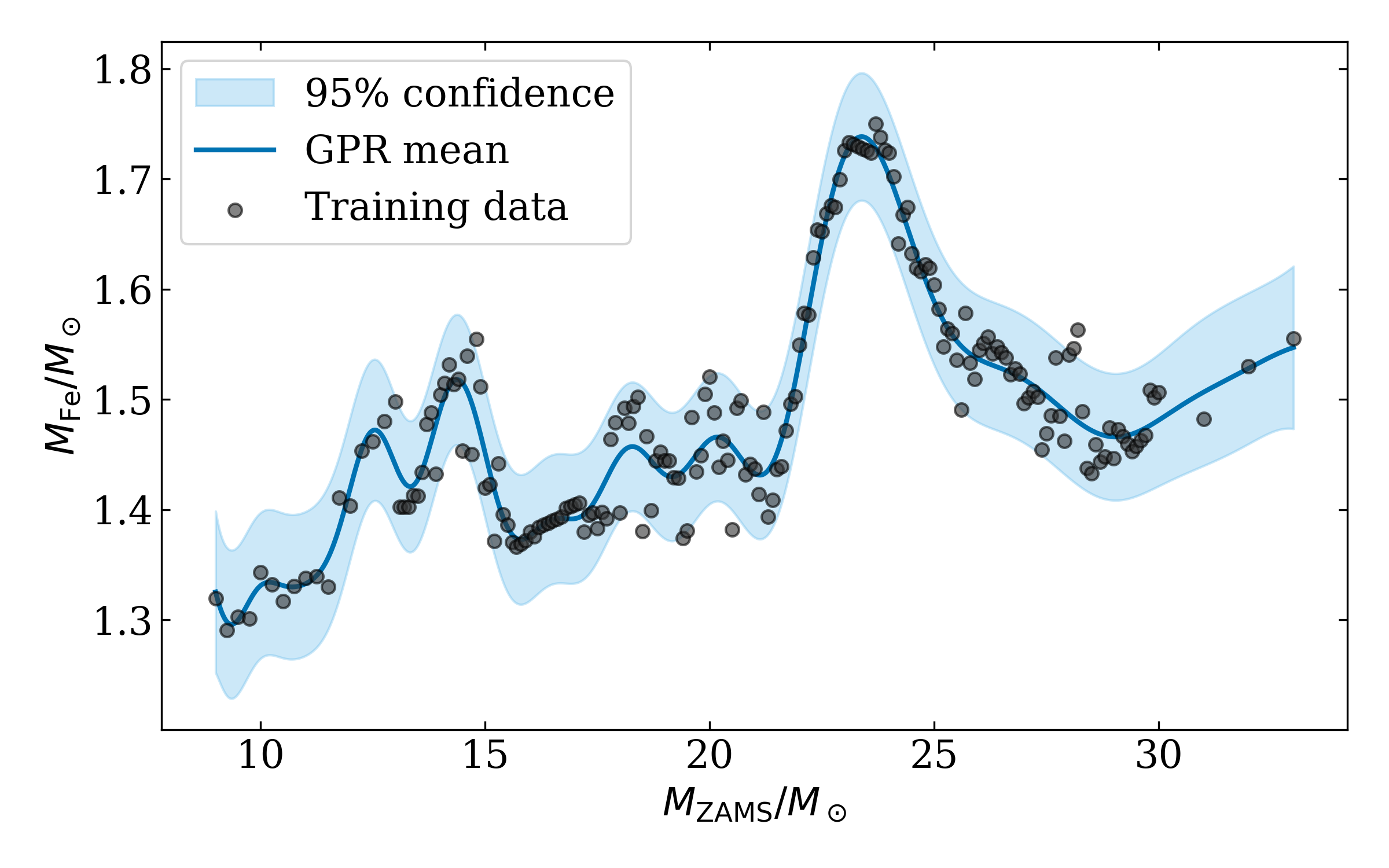}
    \caption{Gaussian process regression fit to the KEPLER grid for the iron core mass $M_{\rm{Fe}}$ as a function of zero-age main sequence mass $M_{\rm{ZAMS}}$. The solid line shows the GP predictive mean, while the shaded region denotes the $95\%$ confidence interval. Black points indicate the 187 simulations used as training data. The GP accurately captures the underlying trends without introducing spurious high-frequency structure, and the majority of the training data lie within the confidence band, indicating well-calibrated predictive uncertainties.}
    \label{fig:fe-core-mass-kepler-fit}
\end{figure}

Throughout this section, we focus specifically on the results for the iron core mass $M_{\rm{Fe}}$, determined by the $^{56}$Fe mass fraction, for the sake of brevity. Analogous results for the other output parameters considered in our analysis can be found in Appendix~\ref{subapp:gpr-for-stellar-parameter-prediction-additional-output-parameters}.  Figure~\ref{fig:fe-core-mass-kepler-fit} displays the results of GP regression on the KEPLER grid iron core mass as a function of $M_{\rm{ZAMS}}$. The mean squared error (MSE) is given by 

\begin{equation}
    \text{MSE} = \frac{1}{N} \sum_{i = 1}^{N} \pa{ y_i - \hat{y}_i }^2,
\end{equation}
with $\hat{y}_i$ the GP predictive mean for the $i$th sample, $y_i$ the true value of the $i$th sample, and $N$ the total number of samples. In this instance, we train on the full KEPLER dataset of $N = 187$ simulations. The MSE for the fit in Figure~\ref{fig:fe-core-mass-kepler-fit} is $6.9 \times 10^{-4}$, indicating strong predictive performance on the training set. To assess the accuracy of the predictive uncertainties (light blue band), we conduct an uncertainty-calibration test. For a confidence level of $\alpha$, we first compute the normalized errors
\begin{equation}
    z_i = \frac{y_i - \hat{y}_i}{\sigma_i},
\end{equation}
where $\sigma_i$ is the predictive uncertainty, and then compute the coverage $\mathcal{C}(\alpha)$,
\begin{equation}
\label{eq:coverage}
    \mathcal{C}(\alpha) = \frac{\text{number of samples with } \ab{z_i} \leq \Phi^{-1}\pa{\frac{\alpha}{2} + \frac{1}{2}}}{N},
\end{equation}
where $\Phi$ is the CDF of the standard normal distribution. This metric assesses whether or not a particular uncertainty level $\alpha$ is accurately captured by the predictive uncertainties provided by the GP --- in other words, $\mathcal{C}(\alpha)$ measures the fraction of observations that lie within the $\alpha$ percentile. A GP's predictive uncertainties are therefore accurate if $\mathcal{C}(\alpha) \approx \alpha$. The results of computing $\mathcal{C}(\alpha)$ for the 0.68, 0.95, and 0.99 percentiles are shown in Table.~\ref{tbl:uncertainty}, which shows that indeed $\mathcal{C}(\alpha) \approx \alpha$. We note that there is some discrepancy for $\mathcal{C}(\alpha=0.68)$, where the GP uncertainty includes more samples than expected. This greater-than-expected coverage provides a conservative overestimate of the uncertainty for lower $\alpha$ and is thus favorable.

\renewcommand{\arraystretch}{1.1}
\begin{table}
\centering
\begin{tabular}{||c|c||c|c||}
\hline
$\alpha$ & $\mathcal{C}(\alpha)$ & $K$ & $R^2$ \\
\hline
0.68000 & 0.77895 & 0 & 0.9159 \\
0.95000 & 0.95789 & 1 & 0.9081 \\
0.99000 & 0.99474 & 2 & 0.9651 \\
{} & {} & 3 & 0.8618 \\
{} & {} & 4 & 0.8490 \\
{} & {} & 5 & 0.9106 \\
\cline{3-4}
{} & {} & Mean & 0.9017 \\
\hline
\end{tabular}
\caption{\textit{Left}: Confidence levels $\alpha$ with associated coverages $\mathcal{C}(\alpha)$ as defined in Equation~\ref{eq:coverage}. Our coverage agrees with the confidence levels except for the $\alpha = 0.68$ case, where our emulators overpredict the uncertainty at the 1-$\sigma$ level. \textit{Right}: K-means fold indices $K$ with their associated $R^2$ values as defined in Equation~\ref{eq:r2}. The average $R^2$ value is shown at the bottom, with a value of $\sim 0.90$ suggesting strong generalization across different folds.}
\label{tbl:uncertainty}
\end{table}

\begin{figure*}[htbp!]
  \centering
  \includegraphics[width=\linewidth]{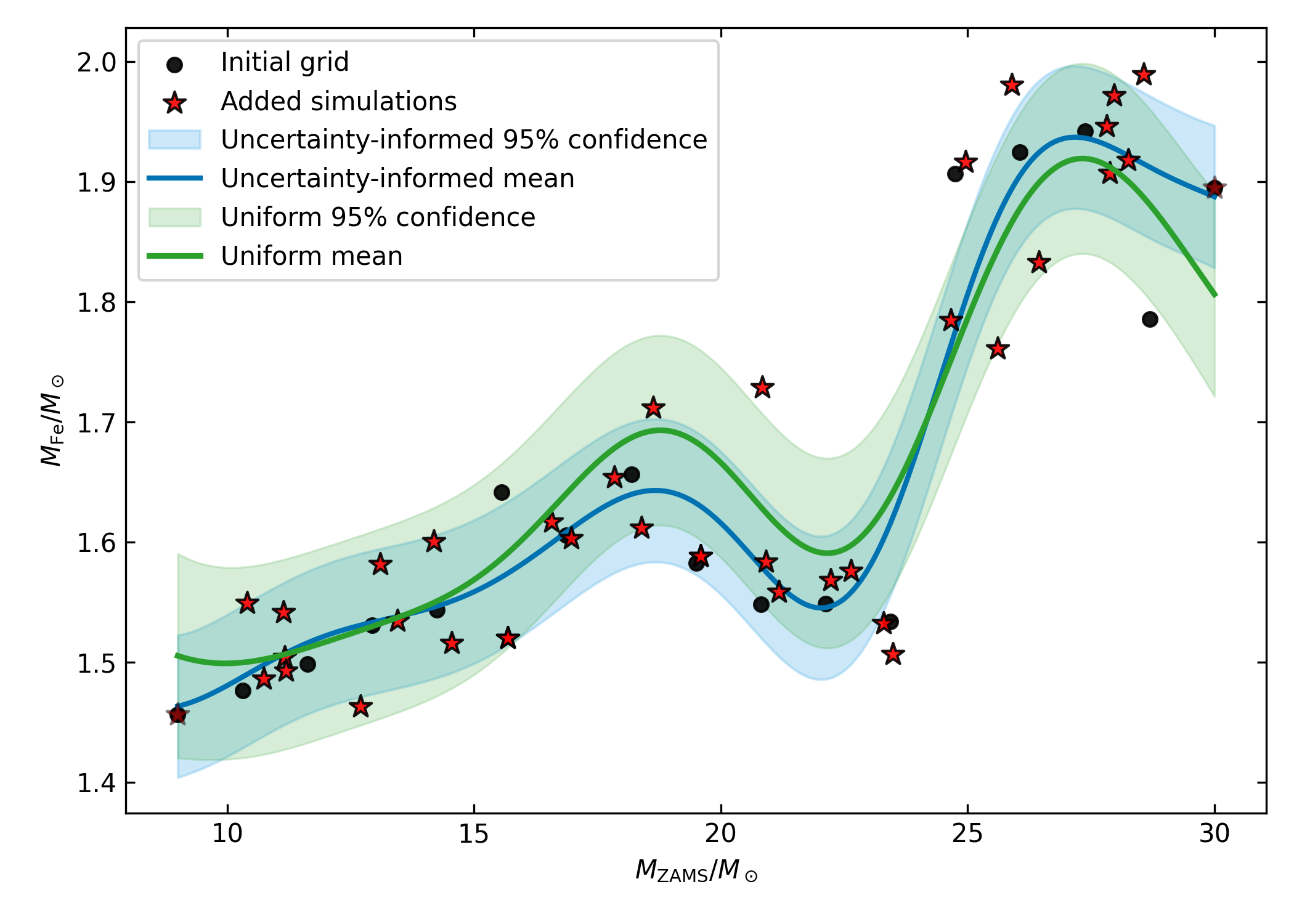}
  \caption{Comparison of Gaussian process regression fits obtained from uncertainty-informed and uniformly spaced simulation grids for the iron core mass $M_{\rm{Fe}}$ as a function of $M_{\rm{ZAMS}}$. Black points denote the initial grid of 17 simulations, while red stars indicate the 48 simulations added through the uncertainty-informed placement procedure (many overlap at $M_{\rm{ZAMS}} = [9, 30]$, see Figure~\ref{fig:simulation-placement-histogram}). We do not plot the uniform grid simulations here. The blue curve and shaded region show the predictive mean and $95\%$ confidence interval for the uncertainty-informed grid, while the green equivalents correspond to the uniformly sampled grid. The uncertainty-informed approach yields systematically reduced predictive uncertainty, particularly near the domain boundaries, demonstrating the advantage of targeted simulation placement. Statistical quantities related to each approach are described in Section~\ref{sec:optimized_vs_naive}.}
  \label{fig:triptych}
\end{figure*}

To assess the generalization of our model to samples outside of training data, we conduct K-fold cross-validation. In particular, we divide our training set into $K = 6$ random subsets, hold out one subset, train the GP and then generate GP predictions on the held out subset, iterating over the $K$ folds. We compute the $R^2$ value for each fold using
\begin{equation}
\label{eq:r2}
    R^2 = 1 - \frac{\sum_{i \in \mathcal{F}_K} \pa{y_i - \hat{y}_i}^2}{\sum_{i \in \mathcal{F}_K} \pa{y_i - \bar{y}}^2},
\end{equation}
for $\mathcal{F}_K$ a set containing the indices of the Kth fold, where $\bar{y}$ is the average test output over the Kth fold. With this test, we measure the effectiveness of our model against the baseline of simply predicting the average output. Our results are shown in Table~\ref{tbl:uncertainty}, with $R^2 \approx 1$ indicating effective predictions on the out-of-sample data. We find an average $R^2 = 0.90174$ across the $K$ folds, indicating strong generalization and predictive performance against the baseline. 

\subsection{Optimized vs. Uniform Placement of MESA Simulations}
\label{sec:optimized_vs_naive}

\begin{figure}[htbp!]
    \includegraphics[width=\linewidth]{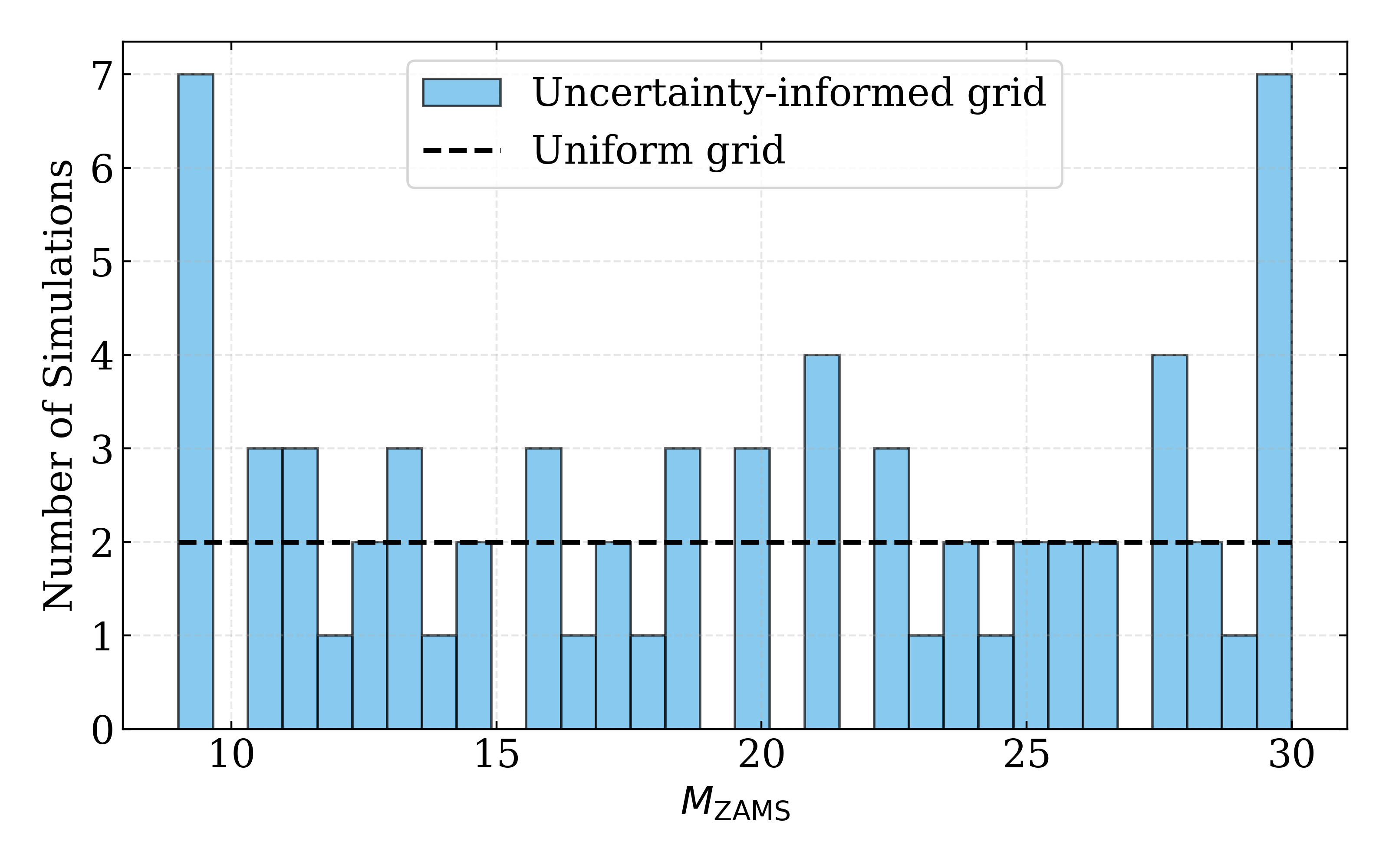}
    \caption{Distribution of simulation placements in $M_{\rm{ZAMS}}$ for the final uncertainty-informed grid. The histogram shows the number of simulations per bin, while the dashed horizontal line indicates the expected count for a uniformly spaced grid. The uncertainty-informed placement exhibits a characteristic U-shaped distribution, with increased sampling near the domain boundaries where predictive uncertainty is largest, reflecting the GP’s tendency to prioritize poorly constrained regions of parameter space.}
    \label{fig:simulation-placement-histogram}
\end{figure}

We evaluate the efficacy of our GP-guided simulation placement method, illustrating in Figure~\ref{fig:triptych} the results of the uncertainty-informed grid-building procedure. The black circles and red stars show the initial and iteratively-placed simulations, respectively, while the solid lines and light shaded regions represent the GP's mean prediction and 95\% credible interval as defined in Equation \eqref{eq:total-uncertainty}. The green lines and shaded regions represent the predictions by the GP trained on a uniformly-placed grid, with the blue equivalents representing the GP trained on the uncertainty-optimized grid. 

Immediately, we notice two key differences between the GP predictions. First, the mean predictions from the two methods deviate from similarity near the $M_{\rm{ZAMS}} = [9, 30]$ limits and for $15 \leq M_{\rm{ZAMS}} \leq 24$. This behavior reinforces the important, though perhaps unsurprising, result that emulator predictions depend not only on simulation number, but on the specific simulation grid. Second, we observe that the uncertainty-informed $95\%$ confidence band is always thinner than the uniform grid's band. Correspondingly, we obtain the average uncertainty values of $U_{\text{blue}} = 0.119$ and $U_{\text{green}} = 0.159$ respectively. The uncertainty-informed confidence band is also much thinner at the domain boundaries (left edge thickness of $0.118$, right edge thickness of $0.118$) than the uniform placement confidence band (left edge thickness of $0.170$, right edge thickness of $0.170$). Note that while the left- and right-boundary values are equal due to rounding, the uncertainty bands are not uniformly wide across $M_{\rm{ZAMS}}$.

We examine the domain edge behavior in Figure~\ref{fig:simulation-placement-histogram} which shows the distribution of simulation placements in $M_{\rm{ZAMS}}$. The distribution of placed simulations reveals a characteristic ``U''-shaped pattern with appreciably more simulations near the domain edges. This behavior arises from the GP's intrinsically higher predictive uncertainty at the boundaries of the training domain, where data are sparse and lack neighboring samples. Consequently, the uncertainty-informed procedure naturally places additional simulations in these poorly constrained regions near the edges to effectively bound the N-dimensional input space hypercube (in this instance, the edges of our 1-D $M_{\rm{ZAMS}}$ space). See also Figure~13 of Ref. \cite{ristic_2020} for additional examples of this phenomenon in a higher dimensional context.

\begin{figure}[tp!]
    \includegraphics[width=\linewidth]{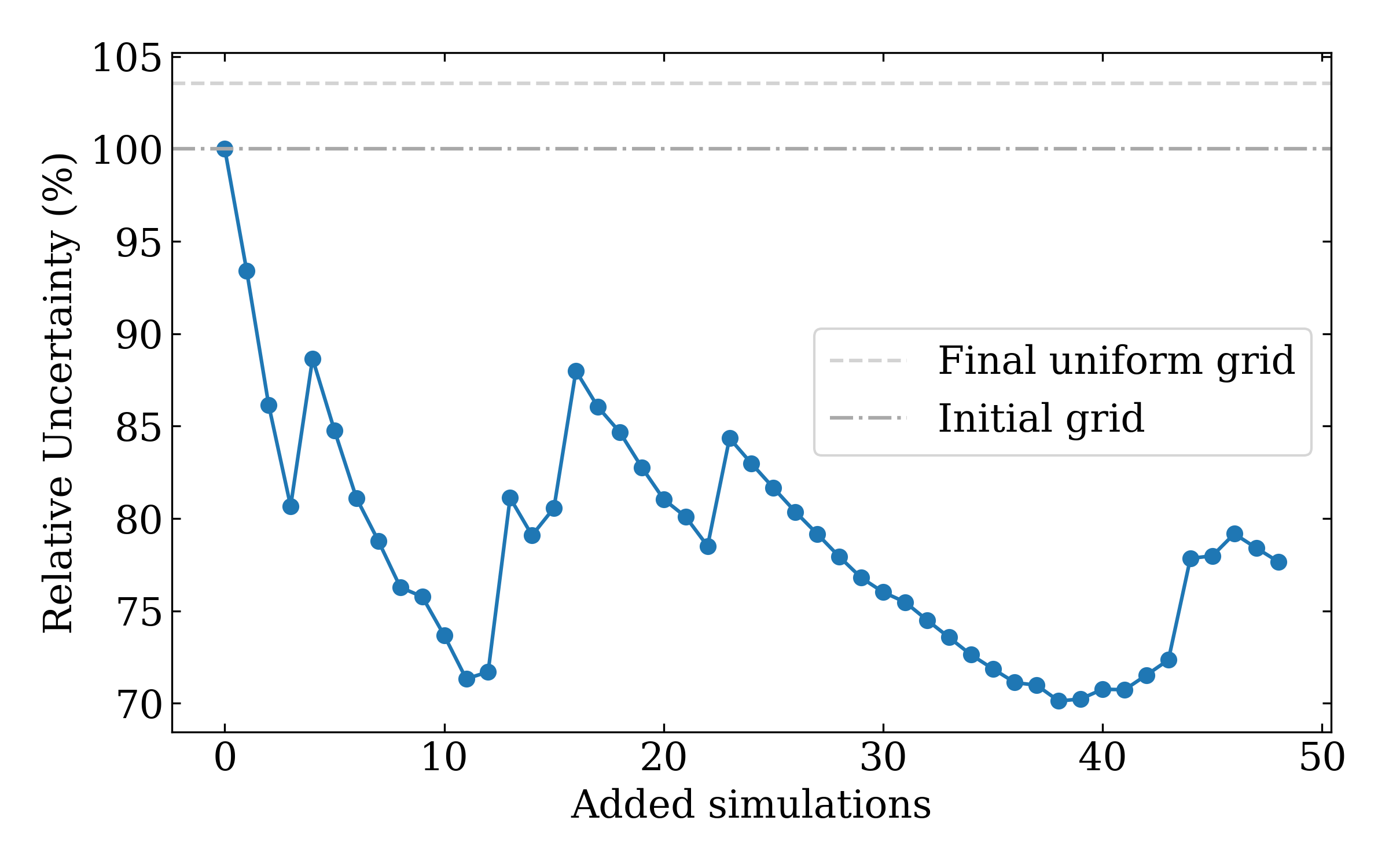}
    \caption{Evolution of the relative uncertainty $U/U_{\rm{initial}}$ as a function of the number of added simulations during the uncertainty-informed grid construction. The solid curve shows the uncertainty reduction achieved by iteratively placing simulations at locations of maximal GP predictive variance. The horizontal dash-dotted (bottom) and dashed (top) lines indicate the uncertainty levels of the initial grid and the final uniformly spaced grid, respectively. The uncertainty-informed procedure yields a substantial overall reduction in predictive uncertainty, especially compared to the uniform placement's increase to over 100\% relative uncertainty. However, the GP uncertainty decrease is not strictly monotonic, reflecting the GP's response to newly resolved structure in the data.}
    \label{fig:uncertainty-curve}
\end{figure}

In Figure~\ref{fig:uncertainty-curve}, we quantitatively track the confidence of the GP, as defined in Equation \eqref{eq:total-uncertainty}, as we build the uncertainty-informed simulation grid. In particular, we plot the evolution of the relative uncertainty which we define as the ratio of the uncertainty of current grid iteration to the uncertainty of the initial grid. The relative uncertainties of the GP trained on the initial and final uniform grids are shown as dashed and dot-dashed horizontal lines, respectively.
Figure~\ref{fig:uncertainty-curve} reveals the unintuitive and remarkably important result that \emph{merely adding more
simulations into the training grid does not generically lead to a reduction in GP prediction uncertainty}. As the dotted line shows, the uncertainty of the GP trained on the uniform grid is \emph{larger} than that of the GP trained on the initial, smaller grid by about 4\%. Meanwhile, the uncertainty of the GP trained on the uncertainty-informed grid is lower by $\sim20-30\%$. We find that the strategic placement of simulations from our uncertainty-informed algorithm leads to a relative uncertainty curve that overall decreases and results in significant improvement over both the initial and uniform grids, keeping in mind that both final grids contain 65 models each. 

\subsection{GP Fits to KEPLER-FLASH Simulation Outputs}

\begin{figure*}[htbp!]
    \centering
    \begin{minipage}[b]{0.48\textwidth}
    \includegraphics[width=\linewidth]{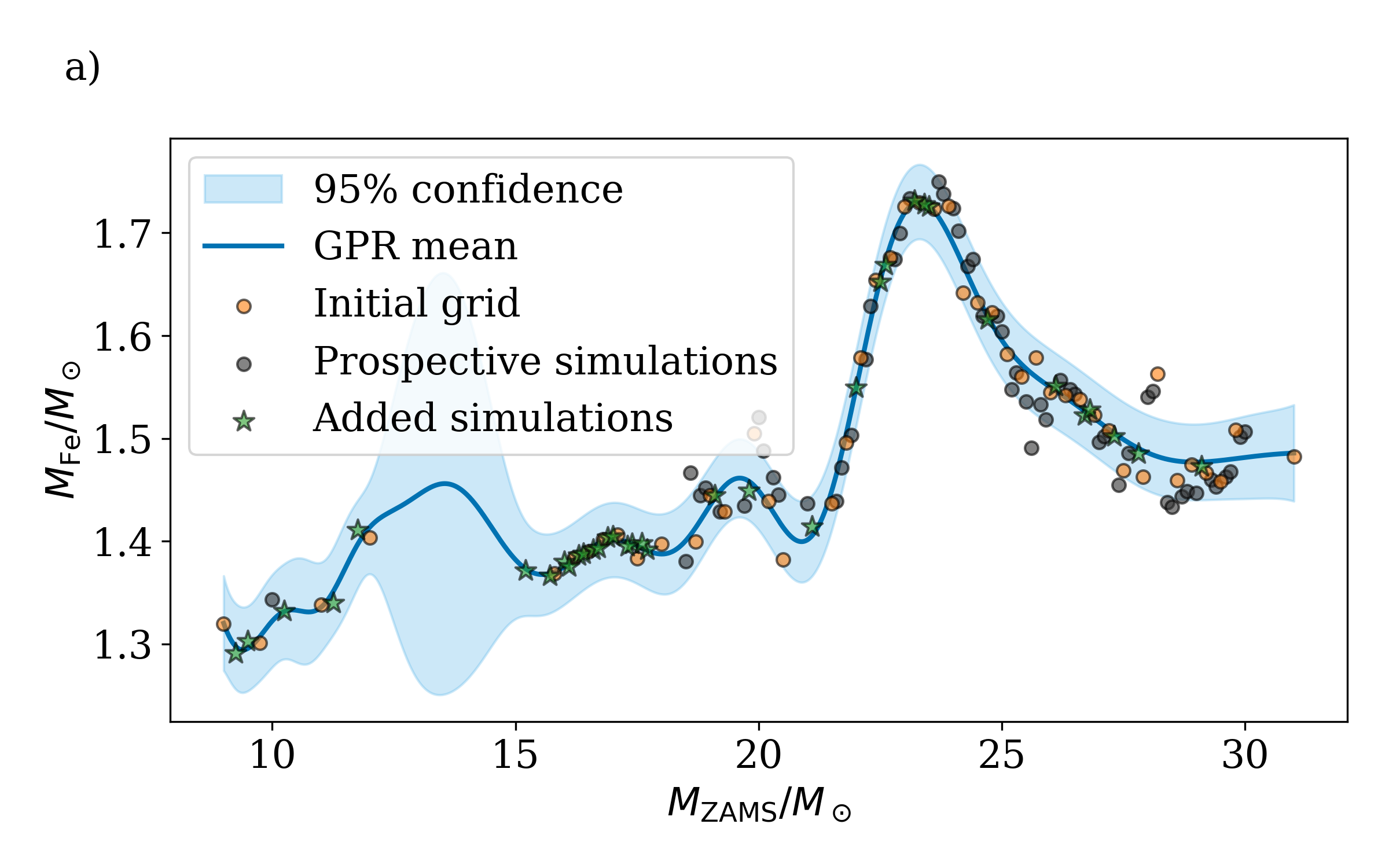}
    \end{minipage}
    \hfill
    \begin{minipage}[b]{0.48\textwidth}
    \includegraphics[width=\linewidth]{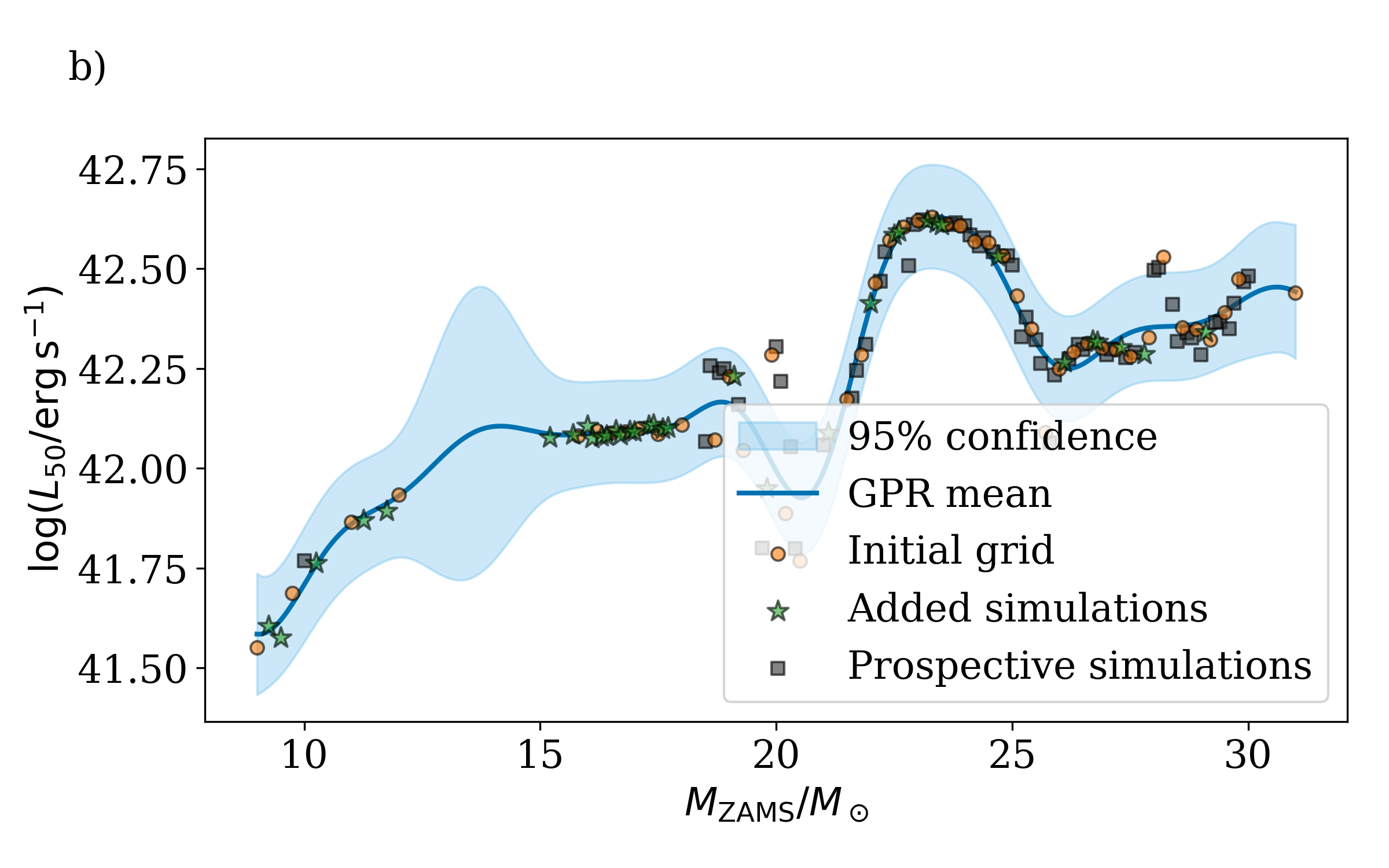}
    \end{minipage}
    
    \begin{minipage}[b]{0.48\textwidth}
    \includegraphics[width=\linewidth]{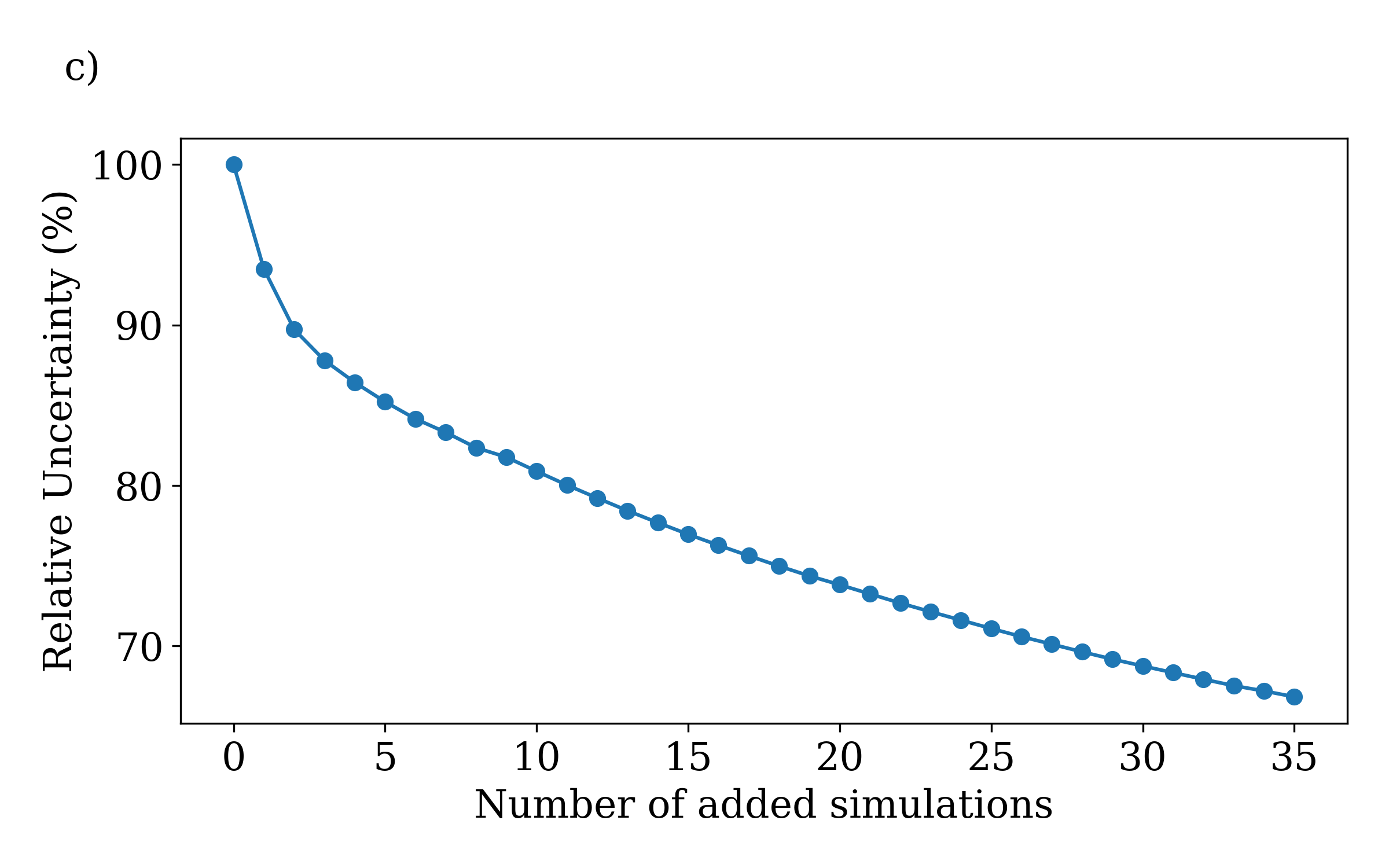}
    \end{minipage}
    \hfill
    \begin{minipage}[b]{0.48\textwidth}
    \includegraphics[width=\linewidth]{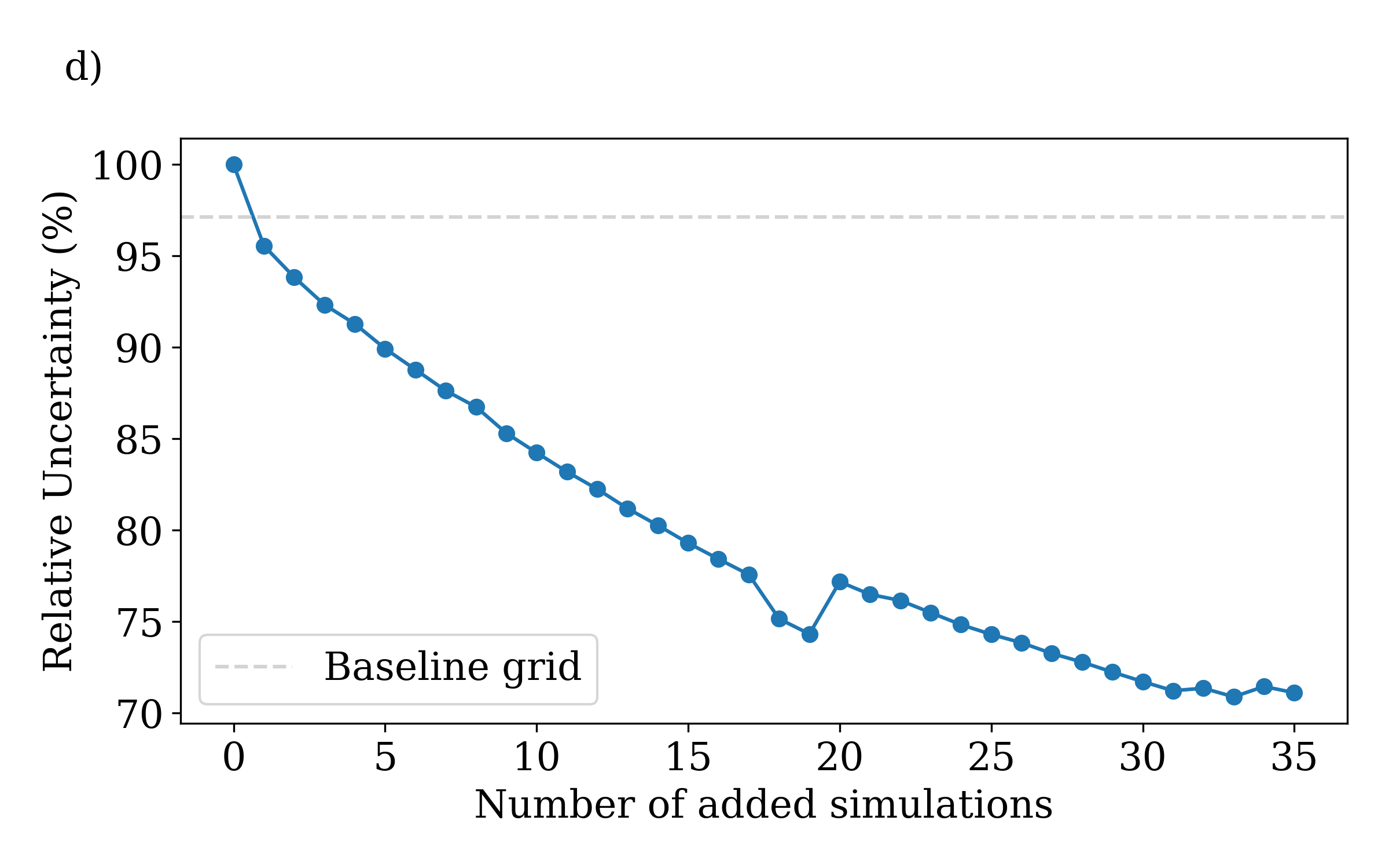}
    \end{minipage}

    \caption{Gaussian process regression results and uncertainty evolution for the KEPLER-FLASH simulation grid. \emph{Top left:} GP fit to the iron core mass $M_{\rm{Fe}}$ as a function of $M_{\rm{ZAMS}}$, showing the predictive mean and $95\%$ confidence interval, along with the initial grid, added simulations, and prospective simulations. \emph{Top right:} Corresponding GP fit for the supernova observable $L_{50}$. \emph{Bottom left:} Evolution of the relative uncertainty for the $M_{\rm{Fe}}$ GP during uncertainty-informed grid construction. \emph{Bottom right:} Evolution of the relative uncertainty for the $L_{50}$ GP evaluated on the same grid, with the dashed line indicating the baseline (randomly selected) grid. Despite optimizing placement using only $M_{\rm{Fe}}$, the resulting grid yields comparable uncertainty reduction for $L_{50}$, demonstrating the effectiveness of upstream, uncertainty-informed sampling for downstream observables. The dashed gray line represents the relative uncertainty of the baseline grid which consists of 22 initial + 35 randomly selected KEPLER-FLASH simulations.}
    \label{fig:four-panel}
\end{figure*}

In this section, we examine how our optimized simulation placement methodology can be used to both generate maximally informative grids and minimize GP prediction variance for downstream science applications. Specifically, we investigate how constructing an optimized grid of stellar evolution simulations can reduce the GP prediction uncertainty of the bolometric plateau luminosity at 50 days ($L_{50}$) for supernovae resulting from these stellar evolution simulations.
As a reminder to the reader, in this setup we do not run new simulations, but instead select a subset of simulations from the prebuilt KEPLER grid.
As described in Section~\ref{sec:gp_flash}, we greedily select a new location from the
KEPLER grid that minimizes the resulting uncertainty in a GP fit to
$M_{\rm{Fe}}$ (not $L_{50}$!). 
The top panels of Figure~\ref{fig:four-panel} show the GP fits to the final grid obtained via our
uncertainty-informed grid construction procedure for both $M_{\rm{Fe}}$ and $L_{50}$ as functions of $M_{\rm{ZAMS}}$.
We obtain roughly the same final relative uncertainties
for both parameters despite carrying out the simulation placement exclusively
based on $M_{\rm{Fe}}$ (we refer the reader to Figure 10 of \cite{barker:2022} and elaborate why this behavior is expected in Section~\ref{sec:disc}). We also note the large ``bubbling'' behavior that occurs between $\sim 13\ M_\odot$
and $\sim 16\ M_\odot$ --- this region of KEPLER simulations leads to black hole
formation without supernovae and thus induces a gap in the KEPLER-FLASH grid.
Thus, the grid construction procedure is forbidden from selecting simulations in
this region and the uncertainty in this region propagates consistently such that subsequent 
inferences are informed of this physical constraint.

In the bottom panels of Figure~\ref{fig:four-panel}, we also explore the evolution of the relative uncertainty as we optimize over the KEPLER grid. The bottom right panel highlights the relative uncertainty of the baseline grid, consisting of 22 initial + 35 randomly selected KEPLER-FLASH simulations, represented by the dashed gray line.
The baseline grid improves the $L_{\rm{50}}$ GP uncertainty by only a factor of a few percent compared to our optimized grid's improvement of $\sim 30\%$.
The significant improvement in uncertainty reduction carried by our method, when compared to the baseline, strongly supports optimized simulation placement for both grid construction and downstream science uncertainty reduction. Achieving this reduction with a subset ($\frac{57}{136}\sim42\%)$ of the total KEPLER-FLASH simulation grid 
is especially appealing for cases where downstream science
applications are dependent on computationally expensive simulation outputs.

\section{Discussion}
\label{sec:disc}

In Figure~\ref{fig:simulation-placement-histogram}, we find that our uncertainty-informed placement procedure exhibits a
characteristic U-shaped sampling pattern that suppresses the intrinsically
larger uncertainties near domain boundaries. This pattern is typically not captured
by common sampling methods such as Latin Hypercube Sampling (LHS), though LHS modifications 
(see, e.g., Ref. \cite{Boschini_2025}) offer promising semi-naive alternatives. 
However, unlike these approaches, we observe that simply adding more simulations
based on a sparsity metric does not guarantee improved GP fidelity, demonstrating
that informed simulation placement is critical to constructing a well-sampled
model grid. While in this study we validate our uncertainty-informed placement procedure using
simplified MESA models, we expect the same qualitative conclusions to hold for
more realistic stellar evolution models.

Similarly, in this work, we perform simulation placement in only one input dimension
($M_{\text{ZAMS}}$); however, more general stellar evolution models depend on a
broader set of initial conditions. ASTRAL is sufficiently flexible to allow
multidimensional input spaces (e.g., with metallicity, rotation), where sparsity
is more severe and active placement is expected to yield even greater efficiency
gains. We anticipate novel behavior in these regimes, such as entropy imbalance,
where placements are preferentially influenced by a single input dimension. This
setup requires balancing the trade-off between pinning down the
now higher-dimensional boundary (e.g., in two dimensions one expects the U-shape
to become a ``bowl'', i.e., a rotated ``U'') and exploring the interior of the
parameter space. We also guide our placement
optimization using a single scalar output as opposed to a multi-objective criterion. In practice, multi-objective
selection will be required for more complex models where multiple outputs contribute to
downstream sensitivity. Such multi-parameter weighting strategies will require
additional stellar evolution physics ansatz. This ansatz can, in part, be developed through
iterative simulation grid placement for which ASTRAL is ideally suited.
While we do not report the results here, we have done preliminary testing of multi-objective
GP fits using the \texttt{GPyTorch} library which supports a ``multitask" vector prediction.
Our analysis, while certainly not exhaustive, suggested that a multi-objective GP
fit was either ill-suited to the KEPLER dataset or required more detailed hyperparameter tuning
than our uniform approach assumed. 

The agreement between $M_{\rm{Fe}}$ and $L_{\rm{50}}$ observed in Figure~\ref{fig:four-panel} may
seem initially surprising, but it follows directly from the connection between progenitor core structure
(i.e. $M_{\rm{Fe}}$) and the energy of the resultant explosion. The fundamental properties of fully
evolved massive stars correlate quite strongly with a number of stellar
evolutionary and explosion properties \citep[see, e.g., Ref.][]{barker:2022}.
For example, they find that the correlation between $M_{\rm{Fe}}$ and $E_{\rm{expl}}$, the supernova
explosion energy, is 0.93, thus explaining the feature similarity in the top panels of Figure~\ref{fig:four-panel}.
This one-to-one relation need not hold generally (as the other correlations in Figure~10 of Ref. \cite{barker:2022} suggest), 
and carrying out the simulation selection on different input and output parameters will generically
lead to different relationships between the resulting uncertainty curves.
Nevertheless, we expect uncertainty-informed placement to outperform uniform placement 
in the context of optimization in generic applications. We also expect that the increased
relative uncertainty observed in the uniform placement is strongly dependent
on both the complexity of features present in the output parameter and the
stochasticity and strength of noise. Future studies, particularly those focused
on more complex parameter space exploration, would benefit from an initial pass at understanding
these features, including with the help of ASTRAL.

\section{Conclusion}
\label{sec:conc}

The ASTRAL framework successfully implements Gaussian process emulation of
stellar evolution models, obtaining strong performance with
reliable predictive means and well-calibrated uncertainties across
a variety of stellar evolution outputs. Furthermore,
ASTRAL proficiently establishes a general framework of uncertainty-informed
simulation placement, the performance of which we demonstrate by (1) running simple MESA
models for optimized grid construction and (2) preferentially selecting a subset of stellar-evolution-informed supernova models
from an fixed simulation library (the KEPLER-FLASH grid).

We also demonstrate ASTRAL's ability to improve GP prediction
uncertainty in downstream science which we demonstrate using supernova
light-curve predictions. Neutrino-radiation-hydrodynamics core collapse supernova simulations represent 
a significantly larger computational expense to optimize as compared to stellar evolution simulations.
By creating  a grid of stellar models that optimally span pre-supernova core
structures we may simultaneously optimize for downstream science while 
minimizing compute time.
ASTRAL represents a dual-optimization of both 
stellar evolutionary and core-collapse supernova objectives, offering greater
and more efficient yields in both computational time and science output.
The ASTRAL code is available on GitHub \cite{data_repo}. %\texttt{https://github.com/astrobarker/astral-gp}.

\section{Acknowledgments}
We thank Floor Broekgaarden and Mathieu Renzo for insightful discussions during the preparation of the manuscript.
We acknowledge the use of the following software: NumPy \citep{numpy}, 
Matplotlib \citep{matplotlib},
SciPy \citep{jones:2001},
\texttt{scikit-learn} \citep{pedregosa2018scikitlearnmachinelearningpython}. \authors{We acknowledge the use of Anthropic's Claude Opus 5 model for generating the analysis and figures in Appendix B only.}
T.G. acknowledges funding support from NSF Career PHY-2340457, NSF AST-2307888, and Simons Foundation award SFI-MPS-BH-00012593-09. 
Los Alamos National Laboratory (LANL) is operated by Triad National Security, LLC, for the National Nuclear Security Administration of U.S. Department of Energy (Contract No. 89233218CNA000001). 
M.~R. acknowledges support from the Information Science \& Technology Institute (ISTI) at LANL as an ISTI Postdoctoral Fellow and acknowledges PECASE award funds.
B.~L.~B. acknowledges support from the Advanced Simulation and Computing Program of LANL as a Metropolis Postdoc Fellow.
The authors acknowledge ChatGPT for its contribution to defining the ASTRAL backronym.
This article is cleared for unlimited release: LA-UR-26-22778.

%%%%%%%%%%%%%%%% APPENDIX %%%%%%%%%%%%%%%%%%%%%%%
\appendix
\section{GP Regression for Stellar Parameter Prediction: Additional Results}
\label{subapp:gpr-for-stellar-parameter-prediction-additional-output-parameters}

Table~\ref{tab:generated} shows the results for GP regression on all of the stellar
output parameters. In general, the average $R^2$ value across the K-folds is
$\approx 1$, except for $T_{\text{PreSN}}$, $R_{\rm{Si}}$, and $M_{\rm{Si}}$ ---
these stellar output parameters exhibit sharp, noisy features, which lead to weaker
mean predictions. However, coverages remain accurate, indicating that the GP can
still quantify its predictive uncertainty, albeit generally, in such cases.

\begin{table}[ht]
\centering
\caption{K-fold cross-validation and coverage for all output parameters. $[M, R,T,L]_{\rm{PreSN}}$ are the mass, radius, temperature, and luminosity of the star before the onset of core collapse; $s_{\rm{c}}, Y_{\rm{e, c}}, \rho_{\rm{c}}$ are the central entropy, central electron fraction, and central density;  $\xi_{M}$ are the compactness parameters defined in Sec.~\ref{sec:sesn}; $M_{\rm{He, CO, Si, Fe}}$ and $R_{\rm{He, CO, Si, Fe}}$ are the masses and radii of the helium, carbon-oxygen, silicon, and iron cores. The $T_{\rm{PreSN}}$ profile is strongly nonlinear as a function of $M_{\rm{ZAMS}}$, but can be naively interpreted as having negative $R^2$ without consideration of more complex features.}
\label{tab:generated}
\begin{tabular}{l c c c c}
\hline
Output Parameter      & Avg.\,$R^2$ & $\mathcal{C}(0.68)$ & $\mathcal{C}(0.95)$ & $\mathcal{C}(0.99)$ \\
\hline
$M_{\text{PreSN}}$    & 0.988 & 0.784 & 0.953 & 0.984 \\
$R_{\text{PreSN}}$    & 0.999 & 0.832 & 0.979 & 0.995 \\
$T_{\text{PreSN}}$    & -0.293 & 0.847 & 0.974 & 0.984 \\
$L_{\text{PreSN}}$    & 0.967 & 0.916 & 0.984 & 1.000 \\
$s_{\rm{c}}$       & 0.910 & 0.816 & 0.942 & 0.989 \\
$Y_{\rm{e, c}}$         & 0.858 & 0.853 & 0.958 & 0.968 \\
$\rho_c$       & 0.876 & 0.821 & 0.947 & 0.989 \\
$\xi_{1.5}$           & 0.821 & 0.774 & 0.947 & 0.974 \\
$\xi_{1.75}$          & 0.807 & 0.758 & 0.958 & 0.984 \\
$\xi_{2.0}$           & 0.893 & 0.816 & 0.926 & 0.995 \\
$\xi_{2.25}$          & 0.921 & 0.800 & 0.937 & 0.974 \\
$\xi_{2.5}$           & 0.925 & 0.795 & 0.937 & 0.974 \\
$\xi_{2.75}$          & 0.934 & 0.800 & 0.942 & 0.974 \\
$\xi_{3.0}$           & 0.944 & 0.805 & 0.942 & 0.974 \\
$M_{\rm{He}}$          & 0.999 & 0.911 & 0.984 & 1.000 \\
$R_{\rm{He}}$        & 0.787 & 0.853 & 0.968 & 0.995 \\
$M_{\rm{CO}}$& 0.999 & 0.942 & 0.984 & 0.995 \\
$R_{\rm{CO}}$& 0.910 & 0.884 & 0.937 & 0.958 \\
$M_{\rm{Si}}$& 0.688 & 0.758 & 0.958 & 0.974 \\
$R_{\rm{Si}}$& 0.105 & 0.853 & 0.942 & 0.963 \\
$M_{\rm{Fe}}$& 0.902 & 0.779 & 0.958 & 0.995 \\
$R_{\rm{Fe}}$& 0.792 & 0.811 & 0.947 & 0.979 \\
\hline
\end{tabular}
\end{table}

\authors{
\section{Scaling Behavior at Higher Dimensions and Extension to Better Suited GP Regression Tools}

More realistic theoretical stellar evolution models will encompass a higher-dimensional parameter space, including, e.g., $M_{\rm{ZAMS}}$, helium abundance $Y$, metallicity $Z$, mass-loss rate $\dot{M}$, initial rotational velocity $v_{\rm{rot}}$, binarity parameters, and more. The standard case of GP regression presented in the main text scales nominally with training set size $N$ as $\mathcal{O}(N^3)$. The scaling behavior stems from the computational cost of calculating the Cholesky decomposition for the lower-triangular factorization of the covariance matrix $\mathbf{K}$. This limitation makes the \texttt{scikit-learn} variation of GP regression intractable as dimensionality $d$, and conversely the training set size $N$, increase in the case of more realistic model grid exploration. However, alternate methodologies exist and allow for some general predictions about expected scaling behaviors within the constraints presented below. 

To examine these approximate scaling behaviors in the limit of increasing $d$ and large $N$, we entertain two deviations from our main text's methodology. 
First, we generate the analytic forms of $M_{\rm{Fe}}$ as a function of up to $d=5$ independent variables presented in Figure~\ref{fig:analytic} which allow us to evaluate the GP's performance at large $N$ without necessitating $N=10,000$ MESA simulations for comparison. The mathematical form of the equations is entirely ad hoc to represent the qualitative features (which should not be interpreted as having any physicality) in $M_{\rm{ZAMS}}$ space and was formulated with the help of Anthropic's Claude Opus 5 model. The input dimensions are not physically meaningful nor representative of stellar evolution modeling parameters; they simply exist as a tool to test GP scaling with a non-trivial function. The structure of the analytic solutions assumes that each dimension's contribution is additive, while coupling two of the parameters to resemble the interdependence of certain quantities such as, for example, $M_{\rm{ZAMS}}$ and $\dot{M}$, while also avoiding a trivial result in which the GP is able to cleanly separate the contributions from each input dimension.

\begin{figure}
\centering
\includegraphics[width=\linewidth]{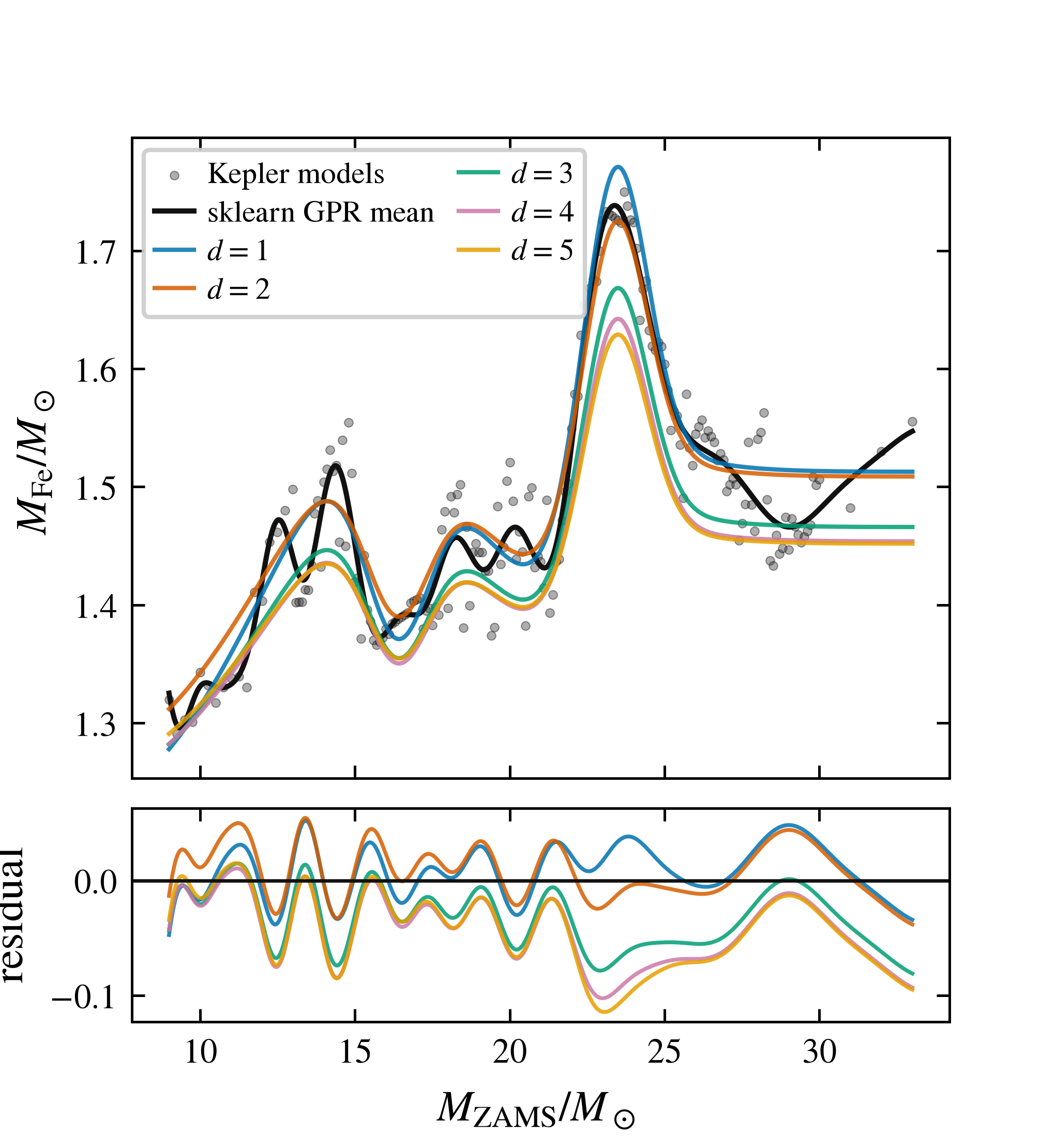}
\caption{Analytic approximations of the mean GP prediction from Figure~\ref{fig:fe-core-mass-kepler-fit} for dimensionality $d$ up to $d=5$. The analytic forms are used to approximate the qualitative features of the iron core mass function space while offering an exact comparison for GP scaling tests in lieu of using thousands of MESA simulations.}
\label{fig:analytic}
\end{figure}

Second, although \texttt{scikit-learn} is entirely well suited to the proof-of-concept problem presented in the main text, it is limited in its ability to scale to higher $d$ and larger $N$ given its use of the exact matrix inversion through Cholesky decomposition. Our scaling study is enabled through a more advanced GP regression library called \texttt{gpytorch} \citep{gpytorch}. The \texttt{gpytorch} framework allows much more flexibility in both the construction of exact and approximate GPs, as well as the training setup flexibility enabling the choice of optimizers, learning rate schedulers, and other training hyperparameters as they appear in the standard \texttt{PyTorch} framework \citep{pytorch}.

The scaling results as a function of $d$ and $N$ are presented in Figure~\ref{fig:scaling}. For each run, we train an exact \texttt{gpytorch} GP for a fixed 200 epochs using the Adam optimizer using 32 bit floating point representation. The 200 epochs are suitable for getting an idea of scaling behavior, but are not guaranteed, and in fact are quite unlikely, to be representative of fully converged GP solutions. Unlike in the main text, we train on an Apple Macbook Pro M5 Max chip, leveraging the metal performance shaders (MPS) to achieve speed-ups over standard CPU training. For the $d=1, N=10,000$ case, our training setup hits the 32 bit noise floor and misrepresents the scaling behavior. Thus, for the $d = 1, N=10,000$ point \emph{and that point only}, we train our GP using the 64 bit floating point datatype on CPUs so as not to misrepresent numerical noise limits as actual scaling behavior. 

The trends in Figure~\ref{fig:scaling} indicate two main features. First, the GP MSE compared to the analytic forms increases as a function of $d$; this result is expected given that $N$ data points cover a smaller volume of the $d$-dimensional hypercube as $d$ increases. Second, the training time appears effectively insensitive to $d$; this comes from \texttt{gpytorch} leveraging iterative conjugate gradient methods to arrive at a solution, reducing training time to $\mathcal{O}(N^2)$. This effect is most readily visible before and after $N=800$, where Cholesky decomposition is used for $N \leq 800$ and conjugate gradients take over for $N > 800$ and the power-law slope of the training time curve relaxes from ${\sim}3$ to ${\sim}2$. We conclude that, for any reasonable amount of MESA simulations (e.g. up to $10,000$ at the upper limit of feasibility), MSE and training time scale to reasonable and efficient values, again with the caveat that our GPs were not trained to convergence. Training setups tailored to individual studies will naturally incur different actual values, but overall we expect the trends to generalize.

\begin{figure}
\includegraphics[width=\linewidth]{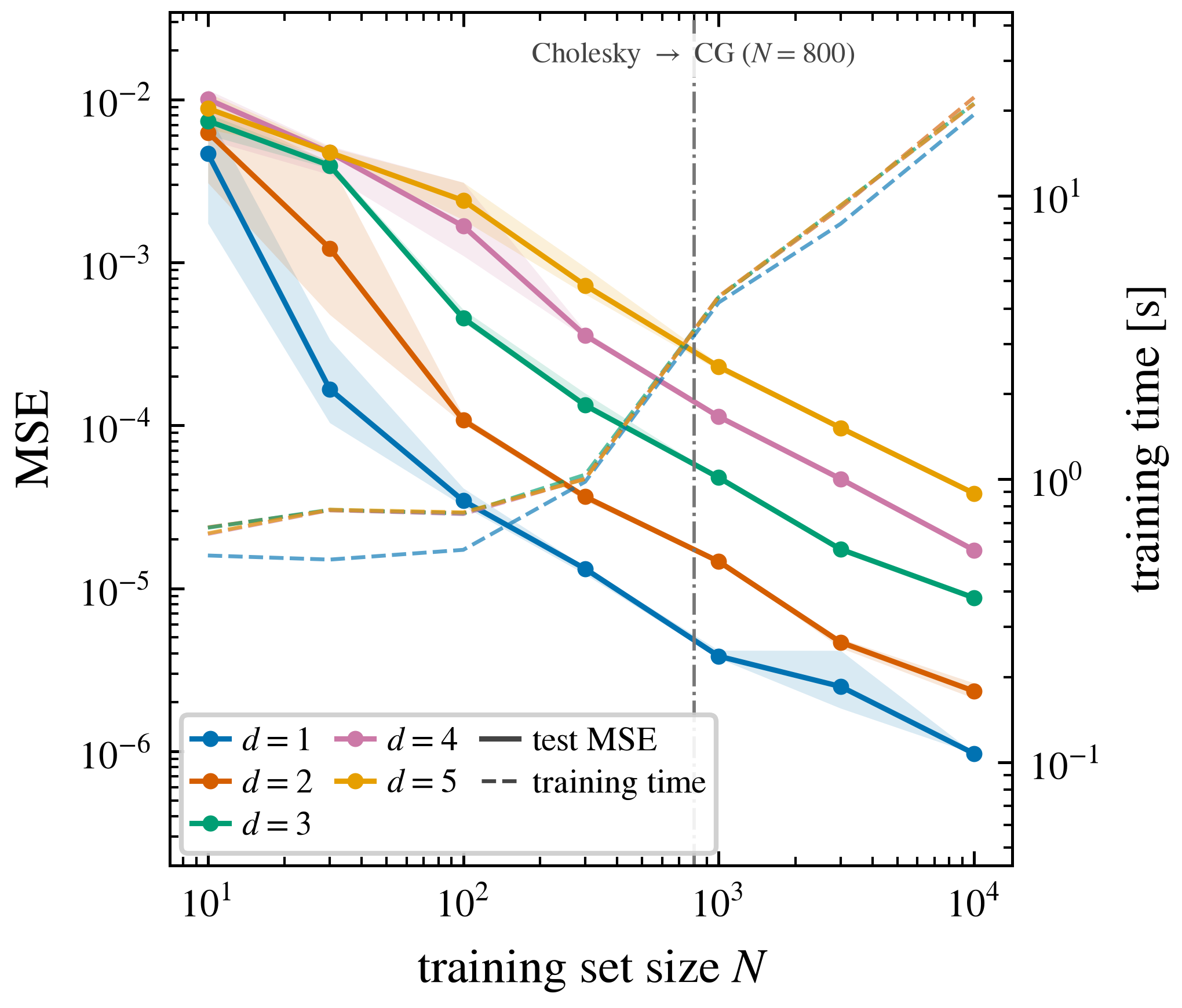}
\caption{GP scaling behavior for a \texttt{gpytorch} exact GP compared to the analytic forms shown in Figure~\ref{fig:analytic} for dimensionality $d$ up to $d=5$ and training set size $N$ up to $N=10,000$. The GP's MSE exhibits dependence on $d$ which is expected given that $N$ points fill up less volume in the $d$-dimensional hypercube as $d$ increases, yielding reduced parameter space coverage. Training time is insensitive to $d$ and dependent solely on $N$, with Cholesky decomposition $(N \leq 800)$ scaling as $\mathcal{O}(N^3)$ and conjugate gradients (CG) scaling as $\mathcal{O}(N^2)$. The change in scaling behavior between these two methods is evident by the training time slope decrease at $N = 1000$.}
\label{fig:scaling}
\end{figure}

Finally, as we consider additional training setups, we must also consider that GP training time may eventually take longer than actual simulation generation. In Figure~\ref{fig:crossover}, we explore the 1-D \texttt{sklearn} and 5-D \texttt{gpytorch} GPs and calculate the number of simulations $N$ at which GP training time will outlast the generation of a single MESA simulation, which we arbitrarily assume takes 30 minutes. We find that, for the 1-D case, the crossover happens at $N \sim 20,000$, which is substantially higher than the expected number of simulations required for convergence. We also approximately recover the $\mathcal{O}(N^3)$ training time expected from standard GP regression techniques. In the 5-D case, we are only able to train GPs for $N \leq 30,000$ before running out-of-memory (OOM). Our calculation thus requires extrapolation based on the training time slope (again nearly recovering the expected $\mathcal{O}(N^2)$) of the last few $N$ and suggests that a GP will take longer than a single MESA simulation for $N > 100,000$. This value is again out of the realm of feasibility for simulation generation. In short, with the suggested \texttt{gpytorch} setups, we expect the GP training time to remain subdominant to the time required to generate a new simulation. While we re-emphasize that our GPs are not trained to convergence, we also note that a sufficiently detailed MESA simulation may take far longer to complete depending on, e.g., nuclear network complexity. Thus, our assumptions underestimate computational costs in both cases, and we expect roughly similar scaling trends to hold in more realistic applications. For any scale of relevance the dominant computational cost remains MESA.

\begin{figure}
\includegraphics[width=\linewidth]{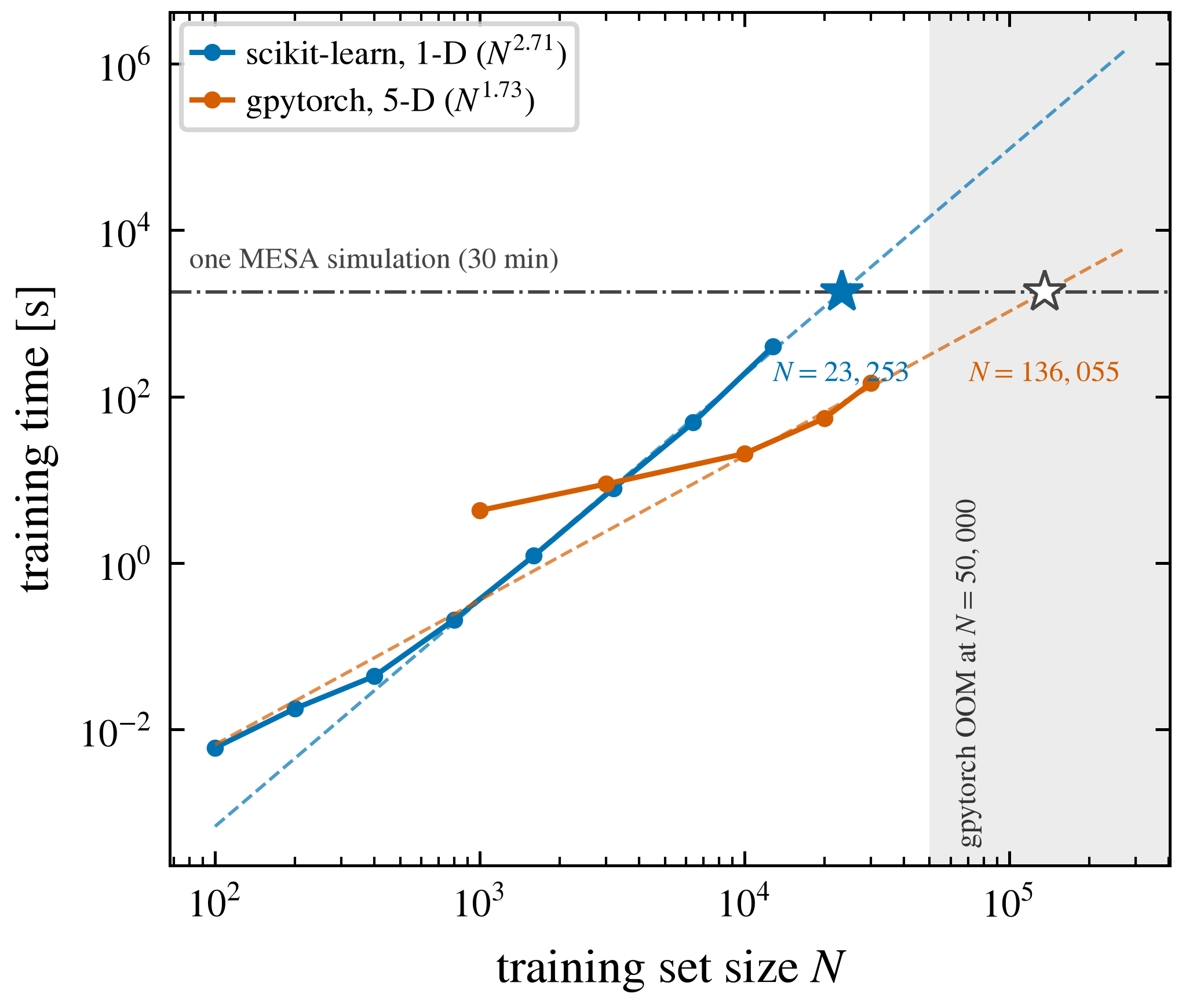}
\caption{Calculation of the GP training time assuming a training set of size $N$. At large enough $N$, the GP training time may exceed the time cost of generating a new MESA simulation, reducing emulator efficacy. In both the \texttt{sklearn} 1-D case and the \texttt{gpytorch} 5-D case, we find that GP training remains subdominant to simulation generation cost in the regime of realistic training set sizes. We note that the longer training time for the \texttt{gpytorch} GP at $N < 10,000$ comes from the additional overhead of CPU-GPU data transfer which is inefficient for small $N$ compared to solely CPU calculation.}
\label{fig:crossover}
\end{figure}

}
\clearpage

%%%%%%%%%%%%%%%%%%%%%% BIB %%%%%%%%%%%%%%%%%%%%%%%%
\bibliography{sample701}{}

\end{document}